\documentstyle[twocolumn,aps,epsf]{revtex}
\def\break#1{\pagebreak \vspace*{#1}}
\def\epsfig#1#2#3#4
         {
         \epsfysize=#2 \vbox{ \hglue#3 \epsfbox[#4]{#1} }
         }
\def\epsfigrot#1#2#3#4
         {
         \epsfxsize=#2
         \setbox\rotbox=\hbox to #2{\epsfbox[#4]{#1}}
         \vbox{\hglue#3 \rotl\rotbox}
         }
\newbox\rotbox
\input rotate

\begin{document}
\draft

\title{  Transferable Tight-Binding Potential for Hydrocarbons }

\author{ Yang Wang and C.H. Mak }

\address{Department of Chemistry,
         University of Southern California,
         Los Angeles, CA 90089--0482}

\date{Revision 2.0: \today}

\maketitle

\widetext
\begin{abstract}

A transferable tight-binding potential has been constructed for
heteroatomic systems containing carbon and hydrogen.  The electronic
degree of freedom is treated explicitly in this potential using a small
set of transferable parameters which has been fitted to small hydrocarbons
and radicals.  Transferability to other higher
hydrocarbons were tested by comparison with {\em ab initio\/} calculations
and experimental data.
The potential can correctly reproduce changes in the electronic
configuration as a function of the local bonding geometry around each
carbon atom.  This type of potential is well suited for computer
simulations of covalently bonded systems in both gas-phase and
condensed-phase systems.

\end{abstract}



\narrowtext

\section{Introduction}

Computer simulation has become an important theoretical
tool in our understanding of complex molecular systems.
The two simulation techniques in
statistical mechanics -- molecular dynamics and Monte Carlo --
commonly used to explore the microscopic origin
of macroscopic behaviors of many-body systems
require as their input
realistic microscopic interactions between the constituents.
In this paper, we describe a general approach for constructing
transferable potentials for heteroatomic systems based on
tight-binding Hamiltonians.
A potential for systems containing carbon and hydrogen
is reported here.  It is part of an ongoing project
to construct potentials for reactive systems of interest
to organic chemistry, biochemistry and materials research.

Systems containing carbon and hydrogen atoms having strong covalent bonds
are interesting because of their central
role in organic chemistry.
Carbon atoms can assume three different hybridizations --
$sp$, $sp^2$ and $sp^3$ -- depending on the local and global chemical
environment, and because of this carbon forms single,
double, triple and conjugated bonds in compounds of many different geometries
with tetrahedral, planar, linear, ring and cage structures.
To understand these systems using computer simulations, it is
necessary to first develop an
interaction
potential for carbon--hydrogen systems which could account for the complexities
of the carbon--carbon and carbon--hydrogen covalent bonds.

Pairwise potentials cannot describe  covalently bonded
hydrocarbons.
A recent effort to construct potentials that can describe reactions
in covalently bonded systems was first made by Tersoff \cite{Tersoff86} who
used an empirical bond--order type potential to describe
primarily silicon--silicon
and other covalent bonds.
This method was later  extended to hydrocarbons
by Brenner\cite{Brenner90} who used  a set of complex functional
forms to parametrize the variation of the bonding geometry on each carbon
atom based on its first and second nearest--neighbor carbon and hydrogen atoms.
The resulting potential involves a large set of
parameters.

\break{1.60in}

In another recent approach, a potential based on the tight-binding
Hamiltonian has been used to describe covalent bonding in homoatomic
systems containing either carbon or silicon
\cite{Chadi78,Harrison,Tomanek86,Goodwin89,Xu92,Tomanek91,Wang92},
as well as
adsorbed hydrogen on silicon
surfaces~\cite{Pandey76,Allan84,Min92}.
This approach treats the electronic
degrees of freedom explicitly
using a small set of transferable parameters.
Chemical bonding results from
filling the single-particle orbitals obtained from
the diagonalization of the tight-binding Hamiltonian.
This alternative approach to constructing interatomic potentials
has many advantages over conventional parametrization schemes.
First, because of its quantum mechanical origin, this potential
has the requisite functional form that is able to describe
bond breakage and bond formation.  Second, only a small set of
parameters is necessary to parametrize a large class of
molecular interactions.  Third, the potential is highly transferable,
i.e. the parametrization is obtained by fitting to only a
very small set of chemical species (typically four or five),
but the potential is extendible to very large systems that
may be substantially more complex.

In this letter, we report a new transferable
minimum-basis tight-binding (TB) potential for hydrocarbons.
We believe this is one of the first successful applications of the
tight-binding potential to describe covalent bonding
and chemical reactions in heteroatomic systems.
Our results on hydrocarbons indicate that using this approach we are able to
generate potential energy surfaces which compare well with
high-level {\em ab initio\/} calculations. This potential is well suited for
dynamical simulations because of
its simplicity and its ability to account for a large variety of bonding
configurations.

\section{method}

Hydrocarbons have strong covalent
two-centered directional bonds.
To describe the hybridization and the directional
bonding character  in hydrocarbons, we adopt a semi-empirical minimum basis
tight-binding Hamiltonian.
The tight-binding Hamiltonian, which considers only the $2s$ and $2p$
valence electrons of carbon and the $1s$ electron in hydrogen, is given by
\begin{equation}
H_{\rm TB} = \sum_{\alpha,i} \epsilon^{i}_{\alpha}a^+_{\alpha,i}a^{
}_{\alpha,i}
      +\sum_{\alpha,\beta,i,j}t^{ij}_{\alpha,\beta}(r_{ij})
       a^+_{\alpha,i}a^{ }_{\beta,j}
\label{eq_tb}
\end{equation}
where $i$ and $j$ label the atoms and $\alpha$ and $\beta$ label the
atomic orbitals. $\epsilon^{i}_{\alpha}$ is the atomic orbital energy of
atom $i$ and orbital $\alpha$.
$t^{ij}_{\alpha,\beta}$ is the hopping matrix
element between atomic orbital $\alpha$ on atom $i$ and $\beta$
on atom $j$.
Eq.~(\ref{eq_tb}) is the conventional tight-binding
Hamiltonian often used for solids \cite{Harrison}.

In our system, there are four types of hopping
matrix elements between carbon atoms: $t^{\rm CC}_{ss\sigma}$,
$t^{\rm CC}_{sp\sigma}$,
$t^{\rm CC}_{pp\sigma}$,
$t^{\rm CC}_{pp\pi}$, and two types of hopping matrix elements
between carbon and hydrogen:
$t^{\rm HC}_{ss\sigma}$,
$t^{\rm HC}_{sp\sigma}$. The detailed forms of the hopping
matrix elements will be discussed later.
Note that since there is no electron--electron interactions in
Eq.~(\ref{eq_tb}), $H_{\rm TB}$ can be exactly diagonalized as a
noninteracting system, yielding
the molecular orbitals.
The electronic configuration is then defined by putting the requisite
number of electrons into the one-electron orbitals subject to the
exclusion principle.

The cohesive energy of the system is defined as
\begin{equation}
E_{\rm coh} =  E_{\rm val}
         + \sum_{i<j} E^{ij}_{\rm core}(r_{ij})
         - \sum_i \ E^i_{\rm atom},
\label{eq_ecoh}
\end{equation}
where $E_{\rm val}$
gives the total energy due to the valence electrons,
the core repulsion energy $E^{ij}_{\rm core}$ comes from the screened ion-ion
interactions between atoms $i$ and $j$, and
the atomic energy $E^i_{\rm atom}$ is the reference energy of the isolated
atom $i$ in the dissociation limit.

The valence energy $E_{\rm val}$ comes from two terms:
\begin{equation}
E_{\rm val} =   \sum_{\gamma} n_{\gamma}\epsilon_{\gamma}
          + \sum_{\gamma} U \delta_{n_{\gamma},2}
\label{eq_eval}
\end{equation}
The first contribution to $E_{\rm val}$ comes from the
molecular orbital energies,
summed over all orbitals resulting from
diagonalizing  the tight-binding Hamiltonian.  $n_{\gamma}$ is the
occupation number of orbital $\gamma$, and
$\epsilon_{\gamma}$ is the single particle  energy of
orbital $\gamma$.
In other words, the first term in Eq.~(\ref{eq_eval})
is simply the contribution of $H_{\rm TB}$ to the total energy.
The second term in Eq.~(\ref{eq_eval}) places a restriction on the electrons
occupying the same molecular orbitals.  This $U$-term describes
empirically
the extra Coulomb repulsion and correlation energy that
arise when two electrons are in the same molecular orbital $\gamma$.
This term contributes only when the molecular orbital is fully
occupied, {i.$\,$e.}, the occupational number $n_\gamma$ is 2.
It is similar in spirit to the $U$-term
in the Hubbard model \cite{Hubbard63}
 and is  included in an attempt to
describe electron correlation in an approximate way.
Because of the $U$-term ($U \geq0$), an energy penalty is paid each
time a pair of electrons (of different spins) go into the same orbital.
Therefore it is possible in some circumstances that total energy is made more
favorable by exciting an electron from a doubly occupied level
to an unoccupied orbital higher in energy
to avoid paying the penalty.
Consequently, the valence energy $E_{\rm val}$ has to be minimized with respect
to the occupation numbers \{$n_{\gamma}$\}.

For the tight-binding Hamiltonian,
we use the
Slater--Koster parameterization scheme\cite{Slater54} for the
electronic hopping matrix elements.
We adopt
scaling forms \cite{Goodwin89}
for the distance
dependence of the hopping matrix elements
as well as for the core repulsive interactions,
\begin{equation}
 t_{\alpha, \beta} (r) = t_{\alpha, \beta}(r_0)
                          \left( \frac{r_0}{r} \right)^{n_a}
                          \exp \left[- n_b \left( \frac{r}{r_{t}} \right)^{n_c}
                                     + n_b \left(
\frac{r_0}{r_{t}}\right)^{n_c}
                               \right]
\label{eq_hopp_ch}
\end{equation}
\begin{equation}
 E_{\rm core} (r)   = E_{\rm core} (r_0)
                      \left( \frac{r_0}{r} \right)^{m_a}
                  \exp \left[ - m_b \left( \frac{r}{r_{c}} \right)^{m_c}
                             + m_b \left( \frac{r_0}{r_{c}} \right)^{m_c}
                      \right]
\label{rep_ch}
\end{equation}
where $r_{t}$ and $r_{c}$ are cutoff distances for the hopping matrix elements
and repulsive interactions,
and the coefficients and power exponents $n_a$, $n_b$, $n_c$, and
$m_a$, $m_b$, $m_c$ determine
the general shape and sharpness of the cutoff functions.
For the same pair of atoms, the reference bond distance $r_0$ is
the same for both the hopping matrix element and the core repulsion.
Unlike the original formulation \cite{Goodwin89}, however,
we do not demand the hopping parameters to have the same
power exponent $n_c$ as the core repulsion $m_c$.

The tight-binding Hamiltonian
explicitly considers the contribution of the
electrons to the total energy of the system.
It is a natural extension of the often used extended H\"uckel model in
organic chemistry~\cite{Huckel}.
In strongly covalently bonded systems,
the molecular orbitals are localized and the chemical bonds
are directional. Consequently, the overlap of the wavefunctions
are mainly along the line connecting two atoms,
and  two-center integrals dominate the
electron-electron interactions, while
three-center and four-center integrals are negligible.
By parameterization of the
hopping matrix elements, we have taken into account
the two-center interactions.
Correlation effects, which are very important in certain circumstances,
are completely missing in the conventional molecular orbital theory.
Some correlation effects have
been reintroduced into our potential via  the Hubbard-like
$U$-term in the Hamiltonian.
Although the $U$-term seems to have been implemented in an {\em ad hoc} manner,
it plays a central
role in determine the occupation of the molecular orbitals.
In particular,
it guarantees that molecules dissociate correctly
without anomalous charges in the
separated atom limit. For example,
a CH radical in the tight-binding Hamiltonian without $U$ will dissociate to
give the unphysical products H$^-$ + C$^+$.
In addition to producing the total energy
of the system, the tight-binding Hamiltonian
also  generates the single particle energy states.
These single-electron energies are useful for
an approximate picture of the
electronic density of states.

There are two parts to the potential: the carbon--hydrogen interactions
and the carbon--carbon interactions.
We will briefly describe how
the parameters of these two parts of the potential were obtained.

\subsection{Carbon-Hydrogen Interactions}

For the carbon--hydrogen interactions, the hopping matrix elements and
repulsion parameters at the reference bond distance $r_0$,
$t^{\rm HC}_{ss\sigma}(r_0)$,
$t^{\rm HC}_{sp\sigma}(r_0)$,
and $E^{\rm HC}_{\rm core}(r_0)$
were first fitted to the experimental energies for
CH, CH$_2$, CH$_3$ and CH$_4$, taken from
Ref.~\cite{Carter88}.
Since the C--H bondlengths in these four CH species are similar, we set
the reference bond distance to $r_0 = 1.09${\AA}.
Care was taken to ensure that the resulting hopping matrix
elements would reproduce a planar equilibrium structure for CH$_3$.
Then the power exponents $n_a$ for the hopping matrix elements,
the power exponent $m_a$ for
the core repulsion and the cutoff functions
were chosen to give the correct experimental equilibrium bond
distance of CH$_4$
as well as the correct breathing
mode frequency of CH$_4$. The resulting parameters are listed in
Table~\ref{tab1}.

\subsection{Carbon-Carbon Interactions}

There are currently available several sets of carbon--carbon
tight-binding parameters \cite{Xu92,Tomanek91,Goodwin91}.
For this work, we have developed a new set.
For the carbon--carbon interactions, the four hopping matrix elements
$t^{\rm CC}_{ss\sigma}(r_0)$,
$t^{\rm CC}_{sp\sigma}(r_0)$,
$t^{\rm CC}_{pp\sigma}(r_0)$,
and $t^{\rm CC}_{pp\pi}(r_0)$,
and the repulsion parameter $E^{\rm CC}_{\rm core}(r_0)$
were fitted to the carbon--carbon bond energies of
C$_2$, C$_2$H$_2$, C$_2$H$_4$, and C$_2$H$_6$
at $r_0=1.312${\AA}. The bond energies of C$_2$ and C$_2$H$_4$
were taken from experiments \cite{Carter88,Hehre86}, whereas the
bond energies
of  C$_2$H$_2$ and C$_2$H$_6$ at
$r_0=1.312${\AA} were deduced from the
respective bond dissociation energies at their
equilibrium distances and the corresponding
force constants of the carbon--carbon bonds.
Then the power exponents for the hopping matrix elements,
the power exponents for the core repulsion
and the cutoff functions were fitted to the equilibrium
carbon--carbon bondlengths and force constants in
C$_2$, C$_2$H$_2$, C$_2$H$_4$, and C$_2$H$_6$.
The resulting carbon--carbon interaction parameters are listed in
Table~\ref{tab1}.

\section{Results}

In this section, we will analyze specific features of our
tight-binding (TB) hydrocarbon potential.
A comparison of the TB potential to
{\em ab initio\/} calculations and experimental data is summarized in
Tables~\ref{tab2} to \ref{tab6}.

\vbox{
\begin{table}
\caption[]{
Parameters for the tight-binding potential.
}
\begin{tabular}{lccc}
$\epsilon^{\rm C}_s$            & --10.290  eV      \\
$\epsilon^{\rm C}_p$            & 0.0   eV          \\
$\epsilon^{\rm H}_s$            & --0.50 eV         \\
U                               & 3.0eV            \\
\tableline
$r^{\rm CH}_0$                  & 1.09 {\AA}      \\
$t^{\rm CH}_{ss\sigma} (r_0)$   & --6.9986 eV     \\
$t^{\rm CH}_{sp\sigma} (r_0)$   & 7.390 eV        \\
$E^{\rm CH}_{core} (r_0)$        & 10.8647 eV       \\
$n_{a}^{\rm CH} $
  for $t^{\rm CC}_{ss\sigma}$   & 1.970          \\
$n_{a}^{\rm CH} $
  for $t^{\rm CC}_{sp\sigma}$   & 1.603          \\
$m_{a}^{\rm CH}$                    & 3.100         \\
$r^{\rm CH}_{t}$                  & 2.0 {\AA}      \\
$n^{\rm CH}_{c}$                  &  9.0          \\
$n^{\rm CH}_{b}$
     for $t^{\rm CH}_{ss\sigma}$ & 1.970      \\
$n^{\rm CH}_{b}$
     for $t^{\rm CH}_{sp\sigma}$ & 1.603      \\
$r^{\rm CH}_{c}$                  & 1.90 {\AA}      \\
$m^{\rm CH}_{c}$                  & 10.0          \\
$m^{\rm CH}_{b}$                  &  3.100        \\
\tableline
$r^{\rm CC}_0$                  &  1.312  {\AA}  \\
$t^{\rm CC}_{ss\sigma} (r_0)$   & --8.42256  eV \\
$t^{\rm CC}_{sp\sigma} (r_0)$   & 8.08162   eV  \\
$t^{\rm CC}_{pp\sigma} (r_0)$   & 7.75792   eV  \\
$t^{\rm CC}_{pp\pi}    (r_0)$   & --3.67510 eV  \\
$E^{\rm CC}_{core} (r_0)$        & 22.68939 eV     \\
$n^{\rm CC}_{a}$
    for $t^{\rm CC}_{ss\sigma}$ & 1.29827      \\
$n^{\rm CC}_{a}$
    for $t^{\rm CC}_{sp\sigma}$ & 0.99055     \\
$n^{\rm CC}_{a}$
    for $t^{\rm CC}_{pp\sigma}$ & 1.01545     \\
$n^{\rm CC}_{a}$
    for $t^{\rm CC}_{pp\pi}$    & 1.82460      \\
$m^{\rm CC}_{a}$                    & 2.72405       \\
$r^{\rm CC}_{t}$                  & 2.00  {\AA}   \\
$n^{\rm CC}_{c}$
    for $t^{\rm CC}_{ss\sigma}$,
        $t^{\rm CC}_{sp\sigma}$,
        $t^{\rm CC}_{pp\sigma}$,
    and $t^{\rm CC}_{pp\pi}$    & 5.0          \\
$n^{\rm CC}_{b}$
    for $t^{\rm CC}_{ss\sigma}$,
        $t^{\rm CC}_{sp\sigma}$,
        $t^{\rm CC}_{pp\sigma}$,
    and $t^{\rm CC}_{pp\pi}$    & 1.0          \\
$r^{\rm CC}_{c}$                  & 1.9  {\AA}   \\
$m^{\rm CC}_{c}$                  & 7.0          \\
$m^{\rm CC}_{b}$                  & 1.0          \\
\end{tabular}
\label{tab1}
\end{table}
}

\subsection{Structures}

In Table~\ref{tab2},  the equilibrium geometries of small
hydrocarbons from the TB potential are
compared to {\em ab initio\/} calculations \cite{Carter88,Hehre86}
and experimental results \cite{Herzberg91}.
Excellent agreement with the experimental values was found.
For the species with only
carbon--hydrogen interactions, i.e. CH$_4$ and CH$_3$,
our semiempirical potential reproduces the correct equilibrium geometries
and the correct carbon--hydrogen bondlengths
to within 0.01~\AA.
For the molecules involving
not only carbon--hydrogen interaction but also
carbon--carbon interactions,
i.e. C$_2$H$_2$, C$_2$H$_4$, C$_2$H$_6$, C$_6$H$_6$,
excellent agreement with experimental results is also achieved.
The carbon--hydrogen bonds in these four hydrocarbons are
only slightly longer compared to experimental data  and
the experimental trend that the carbon--hydrogen bondlength increases from
\vbox{
\widetext
\begin{table}
\caption{
Tight-binding, {\em ab initio\/} and experimental equilibrium geometries
for simple hydrocarbons. All {\em ab initio\/} results are derived from
calculations using a 6-31G** basis set unless otherwise indicated.
All bondlengthes are in \AA.
}
\begin{tabular}{llcdddd}
Molecule
&  Point group
& Feature
& TB
& HF\tablenotemark[1]
& MP4
& Expt.\tablenotemark[2]
\\
\tableline
CH$_3$
&   $D_{3h}$
& r(C--H)
& 1.079
& 1.075
& 1.078
& 1.079
\\
CH$_4$
&   $T_d$
& r(C--H)
& 1.094
& 1.084
& 1.094
& 1.094
\\
C$_2$H$_2$
& $D_{\infty h}$
& r(C--H)
& 1.066
& 1.057
& 1.06
& 1.058
\\
 (triple)
&
& r(C--C)
& 1.183
& 1.186
& 1.22
& 1.208
\\
C$_2$H$_4$
& $D_{2h}$
& r(C--H)
& 1.094
& 1.076
& 1.08
& 1.0836
\\
 (double)
&
& r(C--C)
& 1.341
& 1.316
& 1.34
& 1.337
\\
&
& $\angle$CCH
& 122.8$^{\circ}$
& 121.75$^{\circ}$
& 121.5$^{\circ}$
& 121.25$^{\circ}$
\\
C$_2$H$_6$
& $D_{3d}$
& r(C--H)
& 1.104
& 1.086
& 1.09
& 1.091
\\
 (single)
&
& r(C--C)
& 1.546
& 1.527
& 1.53
& 1.536
\\
&
& $\angle$CCH
& 110.8$^{\circ}$
& 111.2$^{\circ}$
& 111.0$^{\circ}$
& 110.90$^{\circ}$
\\
C$_6$H$_6$
& $D_{6h}$
& r(C--H)
& 1.095
&  1.08\tablenotemark[3]
&  1.09\tablenotemark[3]
&  1.084
\\
(conjugated)
&
& r(C--C)
& 1.428
&  1.38\tablenotemark[3]
&  1.39\tablenotemark[3]
&  1.397
\\
\end{tabular}
\tablenotetext[1]{ Refs. \cite{Carter88} and \cite{Hehre86}. }
\tablenotetext[2]{ Ref. \cite{Herzberg91}. }
\tablenotetext[3]{ 3-21G.}
\label{tab2}
\end{table}
\narrowtext
}
$sp$ to $sp^3$
hybridization is nicely reproduced.
Our TB potential also
reproduces correctly
the trend that
the carbon--carbon bondlength
decreases from C$_2$H$_6$ to C$_2$H$_6$ in the order
single $>$  conjugated $>$ double $>$  triple bonds.
The carbon--carbon bondlengths  in single and double bonds are within
0.01\AA\/ from the experimental values,
while conjugated and triple CC bonds are within 0.03{\AA}\/
from the experimental data.
The bond angles
are also in excellent agreement with experimental measurements.

\subsection{Vibrational Frequencies}

Next we examine the properties of the potential energy surface away from
the minimum.  In Table~\ref{tab3},
the vibrational frequencies are  tabulated and compared to
experimental and {\em ab initio\/} results
\cite{Hehre86,Herzberg91}.
Note that the fitting of the parameters involves only
the C-H and C-C stretching mode frequencies, while all other frequencies
are derived from the TB potential, and they reproduced the
vibrational spectra remarkably well.
Strong anharmonic effects
are observed in small hydrocarbons, and thes
have been well established
theoretically\cite{Surratt77}  and experimentally \cite{Yamada81}.

The vibrational frequencies for CH$_3$ and CH$_4$ are all within 5\%
of the MP2 calculations except for the deformation
mode in CH$_4$ which is within
10\%.
The slightly larger apparent disagreement with the experimental data
on the umbrella
mode of the CH$_3$ is due to strong anharmonic effects.
Figure~\ref{fig1} shows the TB potential energy as a function of
the C--H bondlength for CH$_4$ constrained in the
tetrahedral geometry.
Clearly, the TB predictions agree well with MP4 results
even for large deviations from the equilibrium structure.

In the case of C$_2$H$_2$, the CC stretching frequency
is 7\% larger than the MP2 result.
Other high energy CH modes are in excellent agreement
\break{3.00in}
with the MP2 results.
The two low-energy bending
modes seem stiffer in our empirical potential than the
MP2 results,
but the discrepancy with experimental data can be largely
attributed to anharmonicity.
In the case of C$_2$H$_4$,  all modes agrees well with
the MP2 results, even the
low energy ones.
The CC stretching frequency is correct to within 3\%
compared to the experimental value.
In the case of C$_2$H$_6$, in general all modes agrees well with
the MP2 results.
The CC stretch mode is 10\%
higher than MP2.
The most noticeable discrepancy is the
zero frequency predicted by the TB potential for the torsional motion
because there is no barrier for the torsional motion
in the TB potential for C$_2$H$_6$.
This lack of a torsional barrier is a direct consequence of the
symmetry of C$_2$H$_6$ within the minimum-basis TB Hamiltonian.
The energy barrier in the torsional motion arises from the
steric interaction between the third nearest-neighbor hydrogens and this
component is currently unaccounted for by the present
model.

\subsection{Energetics}

In chemical reactions,
bonds are broken and new ones are reformed.
The reaction energy will depend crucially
on the energetics of different species
involved in the reaction.
In Table~\ref{tab4} and \ref{tab5},
TB predictions
of the energetics involving the breakage of carbon--hydrogen
and carbon--carbon bonds under different situations are compared
to {\em ab initio} calculations
and experimental data \cite{Carter88,Hehre86}.
In Table~\ref{tab4}, the energetics of the sequential
removal of hydrogen atoms
from methane are summarized.
When each hydrogen is being removed,
the remaining carbon--hydrogen bonds are allowed
to relax to the minimum energy configuration.
During the bond breaking process, we start from a stable closed-shell
species CH$_4$. Removal of one hydrogen produces
a planar CH$_3$ radical with one unpaired
electron.
\vbox{
\begin{table}
\caption{
Vibrational frequencies (in cm$^{-1}$) for
                           methane (CH$_4$),
                           the methyl radical (CH$_3$),
                           acetylene (C$_2$H$_2$),
                           ethylene (C$_2$H$_4$),
                      and  ethane (C$_2$H$_6$),
}
\begin{tabular}{lllrrr}
Molecule  & Symmetry & Description
                              &  TB    &  MP2\tablenotemark[1]
                                              &  Expt.\tablenotemark[2] \\
\tableline
CH$_4$ & $a_1$    & s-stretch & 3162   & 3115 & 2917  \\
       & $e$      & d-deform  & 1690   & 1649 & 1534 \\
       & $t_2$    & d-stretch & 3252   & 3257 & 3019 \\
       &          & d-deform  & 1570   & 1418 & 1306 \\
\tableline
CH$_3$ & $a_1$    & s-stretch & 3207   & 3208 & 3004 \\
       & $a_2$    & umbrella mode & 411& 404  & 606  \\
       & $e$      & d-stretch & 3419   & 3398 & 3160 \\
       &          & scissor   & 1552   & 1478 & 1288 \\
\tableline
C$_2$H$_2$ & $\sigma_g^+$ & CH stretch & 3546 & 3593 & 3374 \\
           &              & CC stretch & 2146 & 2006 & 1974 \\
           &              & CH stretch & 3355 & 3516 & 3289 \\
           & $\pi_g$      & CH bend    &  811 & 444  & 612  \\
           & $\pi_u$      & CH bend    &  897 & 783  & 730  \\
\tableline
C$_2$H$_4$ & $a_{1g}$     & CH$_2$ s-stretch & 3272 & 3231 & 3026 \\
           &              & CC stretch       & 1680 & 1724 & 1623 \\
           &              & CH$_2$ scissor   & 1561 & 1425 & 1342  \\
           & $a_u$        & CH$_2$ twist     & 1158 & 1083 & 1023  \\
           & $b_{1g}$     & CH$_2$ a-stretch & 3305 & 3297 & 3103  \\
           &              & CH$_2$ rock      & 1398 & 1265 & 1236 \\
           & $b_{1u}$     & CH$_2$ wag       & 1102 &  980 &  949 \\
           & $b_{2g}$     & CH$_2$ wag       & 1084 &  931 &  943 \\
           & $b_{2u}$     & CH$_2$ a-stretch & 3344 & 3323 & 3106  \\
           &              & CH$_2$ rock      &  899 &  873 &  826 \\
           & $b_{3u}$     & CH$_2$ s-stretch & 3221 & 3222 & 2989 \\
           &              & CH$_2$ scissor   & 1651 & 1523 & 1444 \\
\tableline
C$_2$H$_6$ & $a_{1g}$     & CH$_3$ s-stretch & 3113 & 3086 & 2954 \\
           &              & CH$_3$ s-deform  & 1633 & 1493 & 1388 \\
           &              & CC stretch       & 1157 & 1040 &  995 \\
           & $a_{1u}$     & torsion          &    0 &  452 &  289 \\
           & $a_{2u}$     & CH$_3$ s-stretch & 3148 & 3104 & 2986 \\
           &              & CH$_3$ s-deform  & 1614 & 1494 & 1379 \\
           & $e_g$        & CH$_3$ d-stretch & 3168 & 3228 & 2969 \\
           &              & CH$_3$ d-deform  & 1639 & 1520 & 1468 \\
           &              & CH$_3$ rock      & 1336 & 1264 & 1190 \\
           & $e_u$        & CH$_3$ d-stretch & 3201 & 3215 & 2985 \\
           &              & CH$_3$ d-deform  & 1621 & 1604 & 1469 \\
           &              & CH$_3$ rock      &  911 &  783 &  822 \\
\end{tabular}
\tablenotetext[1]{ 6-31G* MP2 calculations from \cite{Hehre86}. }
\tablenotetext[2]{References \cite{Hehre86} and \cite{Herzberg91}. }
\label{tab3}
\end{table}
}
\vbox{
\begin{table}
\caption{
Calculated and experimental bond energies (in kcal)
for sequential removals of hydrogen from methane.
}
\begin{tabular}{lrrrr}
Theory    & CH$_4$ & CH$_3$ & CH$_2$ & CH   \\
\tableline
TB        & 107    &  119   &  106   & 85   \\
HF\tablenotemark[1]
          & 87     &  88    &  101   & 55   \\
MP2\tablenotemark[1]
          & 109    &  110   &  109   & 73   \\
MP4\tablenotemark[1]
          & 110    &  112   &  107   & 76   \\
Expt.\tablenotemark[2]
          & 112    &  116   &  107   & 84   \\
\end{tabular}
\tablenotetext[1]{ 6-31G** calculations from
                   Refs. \cite{Carter88} and  \cite{Hehre86} }
\tablenotetext[2]{ Ref. \cite{Carter88} }
\label{tab4}
\end{table}
}
\vbox{
\begin{figure}
\epsfigrot{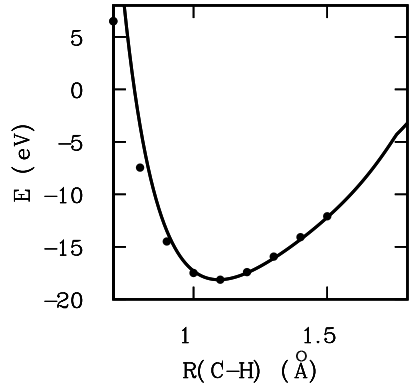}{2.50in}{0.30in}{70 -600 390 -200}
\caption[]{\label{fig1}
    The cohesive energy of methane as a
    function of the carbon--hydrogen bondlength (constrained to
    maintain its T$_d$ symmetry):
    from MP4 6-31G** (circles), from TB (solid line).
    The MP4 results are shifted to the experimental cohesive energy.
}
\end{figure}
}
Removal of another hydrogen atom from CH$_3$ produces a bent
CH$_2$ with two
unpaired electrons. Finally, the removal of a third hydrogen
produces CH with three unpaired electrons.
The sequence of products obtained under sequential removal
of hydrogen from CH$_4$ involves a number of very different
hybridizations, and hence the resulting electronic configurations
and geometries are rather diverse.
Table~\ref{tab4} shows that
our semi-empirical TB potential can
reproduce the energetics quite satisfactorily.
The errors in the sequential carbon--hydrogen dissociation energies are
within 5~kcal from the experimental data, and the
results are much better than {\em ab initio\/} Hartree-Fock calculations.
The order of the carbon--hydrogen bond strengths is also correctly reproduced.

The carbon--carbon bond dissociation energies in C$_2$H$_6$, C$_2$H$_4$
and C$_2$H$_2$ are shown in Table~\ref{tab5}.
In the case of C$_2$H$_6$,
a single carbon--carbon bond is broken and the system dissociates from a
closed-shell structure  into two open-shell methyl radicals. The
double bond in
C$_2$H$_4$ is harder to break than the single bond and the end product
is two open shell CH$_2$ species.  The triple bond in C$_2$H$_2$
contains the most energy among the three.
Our empirical potential
reproduces the three different bond energies excellently.

The diversity of organic compounds relies on the ability of
carbon atoms to assume different hybridizations in order to
form single, double and triple bonds. During chemical reactions,
carbon atoms will rehybridize
from one configuration to another continuously.
A  successful potential should be able to describe such processes
closely.  It will therefore
be a more stringent test for our TB potential to
see if it will reproduce the correct rehybridization behavior
along the reaction path.
\vbox{
\begin{figure}
\epsfig{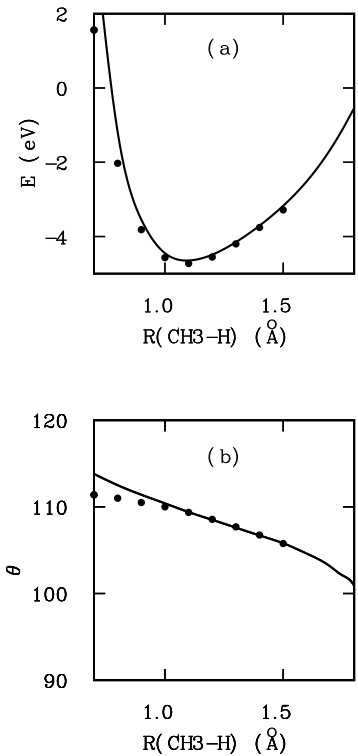}{4.20in}{-0.50in}{-40 -705 0 -100}
\caption[]{
   The calculated
   (a) cohesive energy and
   (b) HCH$_3$ angle
   change due to the variation in the hybridization by
   gradually pulling one hydrogen atom out from methane:
   from MP4 6-31G** (circles), from TB (solid line).
}
\label{fig2}
\end{figure}
}
In Fig.~\ref{fig2}(a), we
show the potential energy curve
for the removal of one hydrogen from the
tetrahedral CH$_4$ to yield a planar CH$_3$.
In the process of pulling one hydrogen out of the
tetrahedron, the carbon atom
reconfigures itself from $sp^3$ to $sp^2$ accordingly
to open up the umbrella formed by the other three
hydrogen atoms
until a planar geometry is attained.
In Figure~\ref{fig2}(b), the HCH$_3$ angle
is plotted as a function of the H--CH$_3$ distance during the
rehybridization of the carbon atom.
The predictions of the TB potential
agree well
with {\em ab initio\/} MP4 calculations.

\subsection{Transferability}

By far, the most significant feature of the TB potential is its high
degree of transferability.
The TB potential has been fitted only to
small molecules and radicals.
In this section, we will demonstrate the
transferability of the TB potential by comparing the TB
predictions to experimental results for very much larger system.

Extended structures can be produced by carbon-carbon bonds.  One
important example is the cage structure of C$_{60}$.
There are 12 pentagons and 20 hexagons in C$_{60}$ and
all 60 atoms are equivalent.
\vbox{
\begin{table}
\caption{
Calculated and experimental carbon--carbon bond energies
(in kcal)
for single, double and triple bonds.
}
\begin{tabular}{ldddddd}
          & {H$_3$C -- CH$_3$}
          & {H$_2$C = CH$_2$}
          & {HC $\equiv$ CH}      \\
\tableline
TB        & 94    & 177   & 235   \\
HF\tablenotemark[1]
          & 69    & 119   & 167   \\
MP2\tablenotemark[1]
          & 100   & 177   & 246   \\
MP4\tablenotemark[1]
          & 98    & 177   & 236   \\
Expt.\tablenotemark[2]
          & 97    & 179   & 236   \\
\end{tabular}
\tablenotetext[1]{ 6-31G**. }
\tablenotetext[2]{ Refs. \cite{Carter88} and \cite{Hehre86}. }
\label{tab5}
\end{table}
}
\vbox{
\begin{table}
\caption{
Comparison of tight-binding predictions for
atomization energies (in eV) of straight-chain alkanes.}
\begin{tabular}{ccc}
n & Experimental\tablenotemark[1] & TB \\
\tableline
1 & 18.22 & 18.13 \\
2 & 30.90 & 31.03 \\
3 & 43.3\tablenotemark[2]  & 43.90  \\
4 & 56.2\tablenotemark[2]  & 56.78  \\
5 & 69.0\tablenotemark[2]  & 69.65  \\
6 & 81.8\tablenotemark[2]  & 82.52  \\
\end{tabular}
\tablenotetext[1]{Ref.~\cite{chains}.}
\tablenotetext[2]{Estimates corrected for zero-point energies.}
\label{tab6}
\end{table}
}
Because of the spherical cage structure,
carbon atoms do not have either pure $sp^3$
or $sp^2$ bonding. The chemical bonds in the pentagons, which are
often called
``single bonds'', are 0.04~\AA\/ longer than the remaining
``double bonds'' in the hexagons.
This difference between the double and single bond distances
is very nicely reproduced by the TB potential,
although the absolute value predicted by the TB potential
for a single bond is 0.05~\AA\/ too long.
The TB potential also predicts
an atomization energy of 7.12~eV per carbon atom, which compares very
favorably with
the experimental result of 7.02~eV per
carbon atom\cite{Steele92,note2}.

Another example of an extended structure produced by carbon-carbon
bonds is the sheet structure of graphite.
The experimental atomization energy of graphite is 7.37~eV per
carbon atom\cite{graphite}.
Ignoring interplanar interactions, our TB potential
predicts 7.40~eV per carbon atom.

The benzene molecule which has conjugated CC bonds
also presents an interesting test for the TB potential.
The experimental atomization energy is 59.56 eV while the TB potential
predicts 59.72 eV.  This yields a C--C bond energy of 145 ~kcal,
which is between the CC single and double bond energies given in
Table~\ref{tab2}.

For the first six straight-chain alkanes C$_n$H$_{2n+2}$,
we have listed the TB predictions for the atomization energy
in Table~\ref{tab6}.  The agreement with experimental
values is almost quantitative.  The TB potential predicts an
average increment of 12.9~eV per additional CH$_2$ unit,
which is very close to the experimental value 12.8~eV.

Finally, we examine one cyclic alkane.
The lowest energy conformation of cyclohexane is the chair form.
Experimentally, the chair form is more stable than the
boat form by 5.5~kcal.
This difference between the chair and the boat forms is due to
steric hindrance which as discussed above
is unaccounted for by the current model.
Therefore, in the present form, the TB potential predicts
essentially no energy difference between them.
The barrier between these two states
is predicted to be about 4.8 kcal/mol
(with the transition state assumed to be the totally
relaxed planar structure).
Experimentally, this barrier is about 10.8 kcal/mol.

\section{Conclusion}

We have described a transferable semi-empirical minimum basis tight-binding
Hamiltonian for treating reactions involving hydrocarbons.
This potential treats the electronic
degrees of freedom explicitly, describing the interatomic interactions
using a small set of transferable parameters.
Chemical bonding results from the diagonalization of the
tight-binding Hamiltonian and then the filling of
the one-particle energy levels with electrons.
This approach can be systematically extended to include more elements across
the periodic table.
Our results on hydrocarbons show that the electronic configuration
depends sensitively on the local bonding configurations
around each atom.
The rehybridization of a carbon atom from $sp^3$ to $sp^2$ under
a change of geometry from a tetrahedral CH$_4$ to a planar
CH$_3$ radical can be followed closely and agrees well with
high level {\em ab initio\/} calculations.
The energetics and topologies of all carbon--carbon bonds ---
single, double, triple and conjugated bonds --- can
be reproduced accurately.
Transferability to substantially larger systems have been demonstrated.
Due to its simplicity and its ability to describe a large variety of
bonding topologies, our approach will be useful for
dynamical simulations of reactive systems
in both the gas phase and the condensed phase.

\acknowledgements

This research is sponsored by the Army Research Office under the
ARO--URI program, Grant Number DAAL03-92-G-0175.
We thank Lance Braswell for his assistance during this
project.



\begin{references}
\bibitem{note2}
The quoted experimental value is determined using an estimate
for the zero-point energy based on an
average vibration frequency of 700~cm$^{-1}$.
\end{references}
\end{document}